\documentclass[12pt]{article}
\usepackage{epsfig}
\begin{document}

\renewcommand{\thefootnote}{\fnsymbol{footnote}}

\begin{center}
{\LARGE \bf Muon astronomy with LVD detector}
\end{center}
\baselineskip=14pt

\begin{center}
\vspace{0.2cm}
{\large \bf LVD Collaboration}\\

\vspace{0.3cm}
M.Aglietta$^{14}$, E.D.Alyea$^{7}$, P.Antonioli$^{1}$,
G.Badino$^{14}$, G.Bari$^{1}$, M.Basile$^{1}$,
V.S.Berezinsky$^{9}$, F.Bersani$^{1}$, M.Bertaina$^{8}$,
R.Bertoni$^{14}$, G.Bruni$^{1}$, G.Cara Romeo$^{1}$,
C.Castagnoli$^{14}$, A.Castellina$^{14}$, A.Chiavassa$^{14}$,
J.A.Chinellato$^{3}$, L.Cifarelli$^{1,\dagger}$, F.Cindolo$^{1}$,
A.Contin$^{1}$, V.L.Dadykin$^{9}$,
L.G.Dos Santos$^{3}$, R.I.Enikeev$^{9}$, W.Fulgione$^{14}$,
P.Galeotti$^{14}$, P.Ghia$^{14}$, P.Giusti$^{1}$, F.Gomez$^{14}$,
R.Granella$^{14}$, F.Grianti$^{1}$, V.I.Gurentsov$^{9}$,
G.Iacobucci$^{1}$, N.Inoue$^{12}$,
E.Kemp$^{3}$, F.F.Khalchukov$^{9}$, E.V.Korolkova$^{9}$
\footnote{e-mail:korolkova@lngs.infn.it},
P.V.Korchaguin$^{9}$, V.B.Korchaguin$^{9}$,
V.A.Kudryavtsev$^{9\dagger\dagger}$,
M.Luvisetto$^{1}$,
A.S.Malguin$^{9}$, T.Massam$^{1}$, N.Mengotti Silva$^{3}$,
C.Morello$^{14}$, R.Nania$^{1}$, G.Navarra$^{14}$,
L.Periale$^{14}$, A.Pesci$^{1}$, P.Picchi$^{14}$,
I.A.Pless$^{8}$, V.G.Ryasny$^{9}$,
O.G.Ryazhskaya$^{9}$, O.Saavedra$^{14}$, K.Saitoh$^{13}$,
G.Sartorelli$^{1}$,
M.Selvi$^{1}$, N.Taborgna$^{5}$, V.P.Talochkin$^{9}$,
G.C.Trinchero$^{14}$, S.Tsuji$^{10}$, A.Turtelli$^{3}$,
P.Vallania$^{14}$, S.Vernetto$^{14}$,
C.Vigorito$^{14}$, L.Votano$^{4}$, T.Wada$^{10}$,
R.Weinstein$^{6}$, M.Widgoff$^{2}$,
V.F.Yakushev$^{9}$, I.Yamamoto$^{11}$,
G.T.Zatsepin$^{9}$, A.Zichichi$^{1}$

\medskip

$^{1}$ {\it University of Bologna and INFN-Bologna, Italy}\\
$^{2}$ {\it Brown University, Providence, USA}\\
$^{3}$ {\it University of Campinas, Campinas, Brazil}\\
$^{4}$ {\it INFN-LNF, Frascati, Italy}\\
$^{5}$ {\it INFN-LNGS, Assergi, Italy}\\
$^{6}$ {\it University of Houston, Houston, USA}\\
$^{7}$ {\it Indiana University, Bloomington, USA}\\
$^{8}$ {\it Massachusetts Institute of Technology, Cambridge, USA}\\
$^{9}$ {\it Institute for Nuclear Research, Russian Academy of
Sciences, Moscow, Russia}\\
$^{10}$ {\it Okayama University, Okayama, Japan}\\
$^{11}$ {\it Okayama University of Science, Okayama, Japan}\\
$^{12}$ {\it Saitama University of Science, Saitama, Japan}\\
$^{13}$ {\it Ashikaga Institute of Technology, Ashikaga, Japan}\\
$^{14}$ {\it University of Torino and INFN-Torino, Italy} \\
\indent {\it Institute of Cosmo-Geophysics, CNR, Torino, Italy}\\
\vspace{0.2cm}
$^{\dagger}$ {\it now at University of Salerno and INFN-Salerno, Italy}\\
$^{\dagger\dagger}$ {\it now at University of Sheffield, UK}\\
\end{center}
\vspace{0.2cm}

\begin{center}
{\large \bf Abstract\\}
\end{center}

We analysed the arrival directions of single muons detected by the
first LVD tower from November, 1994 till January, 1998. 
The moon shadowing effect has been observed.
To search for point sources of high energy photons
we have analysed muons crossing the rock thickness greater than 3, 5 and
7 km w.e.,  which corresponds to the mean
muon energies  1.6, 3.9 and 8.4 TeV at the surface,
respectively. Upper limits on  steady muon fluxes for selected
astrophysical sources for different muon energies are presented.

\pagebreak

\section{Introduction:}

During  recent years major discoveries have been made in  very 
high energy (VHE) $\gamma$-ray astronomy. Ground-based experiments operating 
at TeV energies using atmospheric 
Cherenkov technique have unambigously detected $\gamma$-rays from a handful of 
sources at VHE (for recent review see Ong, 1998). 
Six sources ( Crab Nebula, Mrk501, Mrk421, Vela, SN1006 and PSR B1706-44) 
were observed with 
significance levels in excess of 6 standard deviations above background. Their 
spectra have been measured up to maximum energies 10-50 TeV. $\gamma$-rays from 
these sources can initiate muons with probability of order 1\%. Muons originate 
from decay of pions, kaons and charmed particles 
produced by shower photons and from muon pair 
production by photons. The production of high-energy muons
in gamma-induced showers
in the atmosphere and possible detection of muons underground were
discussed by Kudryavtsev and Ryazhskaya (1985), Stanev et al. (1985),
Stanev (1986), Berezinsky et al. (1988), Halzen et al. (1997).
Since 1985, a number of experiments has looked for a muon excess from known 
sources and so far the old results of NUSEX  (Battistoni et al., 1985) and 
SOUDAN (Marshak et al., 1985) collaborations which detected muon excess 
from Cyg X-3 have not been confirmed by other experiments 
(Ahlen et al., 1993, 
Giglietto et al., 1997, Poirier et al., 1997). 
A new interest for muon astronomy arises 
from the recent success of ground-based $\gamma$-ray astronomy. 

Single muons observed by LVD detector have been used to search for 
a possible flux from
$\gamma$-ray sources discovered by ground-based experiments in the northern 
hemisphere as
well as from some other known sources. Here we present the results of 
such analysis.	   

\section{Detector and Data Analysis:}

The data used for the analysis were
collected with the 1st LVD tower during 22789 hours of live time.
The 1st LVD tower contains 38 identical modules and has dimensions 
13m $\times$ 6.3m $\times$ 12m. 
Each module consists of 8 scintillation counters and 4 layers of limited
streamer tubes (tracking detector) attached to the bottom and to one
vertical side of the metallic supporting structure. Geometric acceptance 
for isotropic flux is
about 1700 m$^2$ sr. A detailed description of 
the detector was given in Aglietta et al. (1992).    
The depth of LVD site (42$^o$27$'$ N and 13$^o$34$'$ E) averaged over the muon 
flux is about 3650 hg/cm$^2$ which corresponds to the median energy of 
vertical muons at  sea level of about 2.2 TeV. LVD detects
muons crossing from 3000 hg/cm$^2$ to more than 12000 hg/cm$^2$ of rock (which
corresponds to the median muon energies at sea level from 1.6 TeV to
40 TeV for conventional atmospheric muons) 
at zenith angles from 0$^o$ to 90 $^o$ (on the average, 
larger depths correspond to higher zenith angles). This allows us to 
analyse muons in different energy ranges. Three muon samples 
have been chosen in our study: 
1) muons crossing all column densities of rock  
(corresponding energy threshold defined as the median surface
energy of conventional
vertical muons which cross the minimal rock thikness of 3 km w.e.,  
$E^{thr}_{\mu}$, is equal to 1.6 TeV), 
2) muons 
crossing rock thickness greater than 5 km w.e. ($E^{thr}_{\mu}$=3.9 TeV) 
and 3) muons crossing 
rock thickness greater than 7 km w.e. ($E^{thr}_{\mu}$=8.4 TeV).

Muon celestial coordinates have been stored in two dimensional map
with a cell size of 1$^o$ in right ascension (R.A.)
and 0.01 in $\sin\delta$ (where $\delta$ is declination).

\section{Results and Discussion:}

We used the shadowing of cosmic rays by the Moon to confirm the 
pointing accuracy of the LVD detector. 
The data used in the search for the shadow
of the Moon included 1.85$\cdot$10$^{6}$ muons. For every muon 
arrival time, R.A. and
$\delta$  of the geocentric apparent position of the center of the Moon has 
been computed taking into account the corrections for parallax. The angle 
between muon direction and the position of the center of the Moon has been 
evaluated.
We simulated the background events  from the experimental 
zenith-azimuthal distribution
of muons and the mean time between two consecutive muons observed by LVD run
by run. Then the angle between background event and Moon position has been 
calculated.
Figure 1 shows the angular density $dN/d\Omega$ as a function 
of the
angular distance from the center of the Moon.  
The observed deficit has a significance of 2.62 standard deviations (s.d.). 
This study confirms 
that the track reconstruction and pointing accuracy have no serious 
systematic errors.

\vspace{-1.5cm}
\begin{figure}[htb]
\begin{center}
\mbox{\epsfig{file=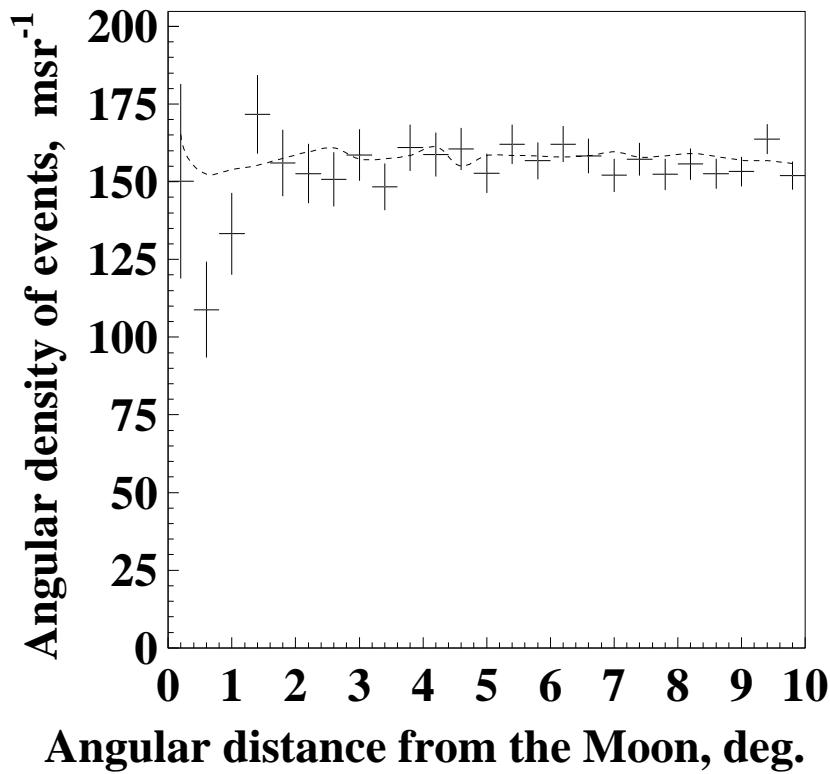,height=12cm}}
\caption
{The angular density $dN/d\Omega$ as a function of the angular distance from
the Moon; the distribution for the simulated background events is shown by
dashed line. }
\end{center}
\end{figure}
 
The distribution of the data versus declination (after summing 
over R.A.) for three 
selected ranges of depth is presented in Figures 2a, 2b, 2c together 
with calculated background of atmospheric muons. The difference in the 
distributions for three depth ranges 
reflects  different mountain structure for these regions at LVD site. 
Figure 2d shows the distribution of the muon flux versus R.A. 
(summed over declination). The calculated background fits data for 
three analysed depth ranges rather well.

\begin{figure} [htb]
\vspace{-2.0cm}
\mbox{\epsfig{file=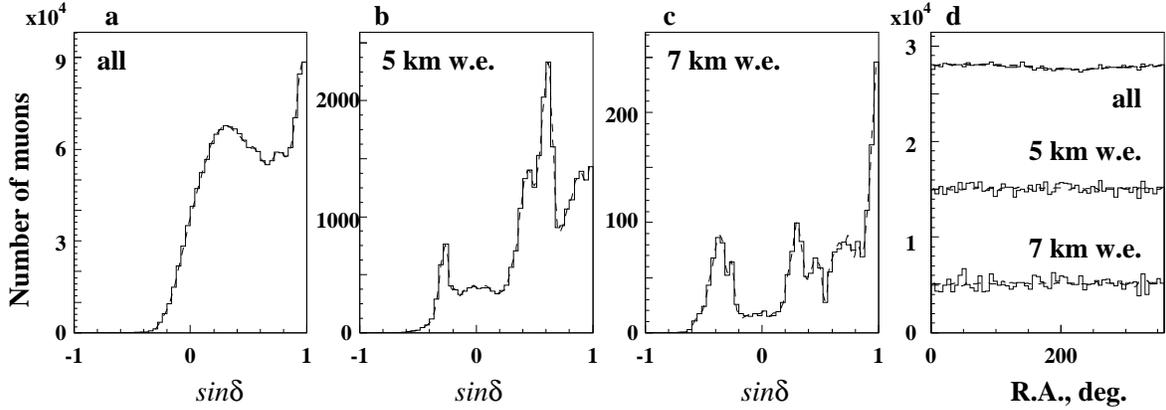,width=16cm}}
\vspace{-0.9cm}
\caption
{Distribution  of muon events vs $\sin\delta$ (a,b,c) and R.A. 
(d) for three depth ranges analysed. 
The histograms are experimental data,
dashed curves show simulated backgrounds from atmospheric muons initiated
in hadronic showers. The histograms for the second and the third depth ranges 
in Figure 2d
were multiplied by factors of 15 and 66, respectively. }
\end{figure}
\vspace{0.1cm}
To test the presence of a significant muon excess above the background 
from any angular cell in the sky we 
used cells
of equal solid angles with a width of $3^o$ in R.A. and 
0.04 in $\sin\delta$ which corresponds approximately to the solid angle 
of a cone with a half angle of 1.5$^o$.
The deviation from the mean was computed using the
Gaussian statistics $\frac{n_{exp}-n_{mc}}{\sqrt{n_{mc}}}$, where
$n_{exp}$ is a number of muons observed in the experiment and $n_{mc}$ is the  
simulated background from atmospheric muons. No cell
with an excess of more than 3.5 s.d. has been found for the first depth range. 
The Gaussian fit gives $\chi^{2}/Dof=0.86$. 
For the second range we found two cells with excesses of 3.72 s.d. and 3.84 s.d.
and worse Gaussian fit with $\chi^{2}/Dof=1.41$.
We shifted the cells by $1^o$ in R.A. and 0.01 in $\sin\delta$, 
repeated this procedure  and did not find any excess more than
3.5 s.d in the overlapping bins. To have better 
statistics for the third depth range the cells have been enlarged up to  
$10^o$ in R.A. and 0.1 in $\sin\delta$. Two bins with excesses of 
4.19 s.d. and 3.83 s.d. were  found but disappeared after the cells
were shifted in the same way as for the second depth range. 
Gaussian fit for this range gives $\chi^{2}/Dof=1.29$. 
We conclude that we have not found significant excess of muons over
background from any cell on the
sky in the selected depth ranges. Figure 3 shows the results of this study.

\vspace{-1.2cm}
\begin{figure} [htb]
\mbox{\epsfig{file=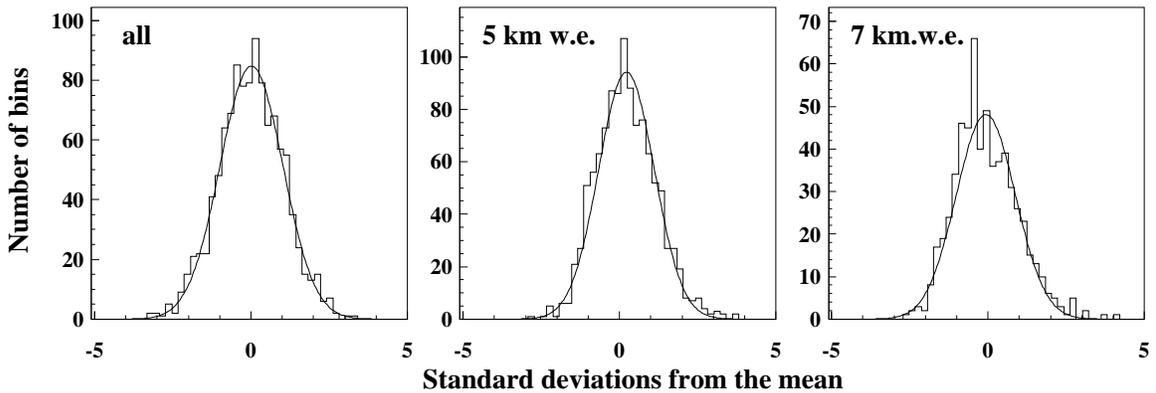,width=16cm}}
\vspace{-0.7cm}
\caption
{Distribution of deviations from the mean value of muon flux and Gaussian fit.}
\label{fig3}
\end{figure}

We have also made a search for a possible flux in narrow cones ($1.5^o$ half 
angle) around the position of the sources  
observed in $\gamma$-ray ground-based experiments
(visible in the northern hemisphere)
and some other sources
which have drawn attention of underground experiments during the last decade. 
Again the data from three depth ranges were considered. To obtain an upper 
limit (95\% C.L.) on the flux from a source we used the following formula:
\begin{equation}
F = \frac {n_{\mu}}{f\cdot\langle
\epsilon\cdot A\rangle\cdot k\cdot T}
\end{equation}
where $n_{\mu}$ is the upper limit on the number of muons calculated 
according to the procedure
given in (Helene, 1983), $f$=0.9 is correction factor to calculate the muons 
scattered out of $1.5^o$ half angle cone, this factor is estimated taking
into account muon deflection in the rock convolved with the simulated detector 
response function for single muons, 
$\langle\epsilon\cdot A\rangle$ is the weighted average of the product
of efficiency of muon detection and reconstruction times the area of the
cross--section of the detector perpendicular to the muon track,
it depends on source position and the depth range, for Cyg X--3 and the first
depth range $\langle\epsilon\cdot A\rangle$=$84\cdot 10^{4}$ cm$^{2}$, 
$k$ takes into account
the time when the source is visible and T is the exposure time.
The results are presented in Table 1.


\begin{table}[htb]
\caption{Steady flux limits (95\% C.L.) from selected sources.
$N_{all}$ , $N_{5}$ and N$_{7}$  
show the number of muons observed
(simulated) in the corresponding cones for three depth ranges,  
$N_{7}^{cut}$ is the same value after depth cut (see text for details), 
$F_{all}$,
$F_{5}$ and $F_{7}$ are the flux upper limits (95\% C.L.) multiplied
by a factor $10^{13}$, cm$^{-2}$s$^{-1}$}
\vspace{0.2cm}
\begin{center}
\begin{tabular}{|c|c|c|c|c|c|c|c|}
\hline
Source & $N_{all}$ & $N_{5}$ & $N_{7}$ & $N_{7}^{cut}$ & $F_{all}$ &
$F_{5}$ & $F_{7}$ \\
\hline
Cyg X--3 & 474(506) & 26(32) & 1(1.4) & 1(0.9) & 11.7 & 3.70 & 1.17 \\
Her X--1 & 528(544) & 44(34) & 1(1.3) & 1(0.9) & 14.2 & 7.66 & 1.28 \\
Crab Nebula & 618(620) & 19(17) & 1(0.6) & 1(0.5) & 19.5 & 3.51 & 1.46 \\
SS433 & 499(465) & 5(8) & 0(0.2) & 0(0.2) & 29.5& 2.18 & 1.29 \\
3C273 & 429 (405)& 5(7) & 0(0.4) & 0(0.4) & 28.3& 2.71 & 1.46 \\
Geminga & 653(629) & 14(10) & 3(1.5) & 3(1.4) & 25.2 & 4.37 & 2.33 \\
Mrk 421 & 565(528) & 36(44) & 3(1.6) & 1(1.0) & 21.5 & 4.79 & 1.28 \\
Mrk 501 & 497(503) & 47(38) & 6(1.4) & 4(1.0) & 13.5 & 7.44 & 2.44 \\
\hline
\end{tabular}
\end{center}
\end{table}

No statistically significant enhancement (more than 1.5 s.d.)
of observed muons above calculated background of atmospheric muons 
has been found from 
any source for all-depth range and for slant depths 
greater than 5 km w.e. For the range of slant depths more than 7 km w.e. 
6 muon events against the background of 1.4 events were observed from
the angular cell which includes
Mrk 501 (this corresponds to the probability of 0.0031). However the excess can
be connected with the complicated mountain structure at these depths. 
To test this hypothesis a special 
depth cut has been applied both for the observed and simulated events. 
For every muon we
calculated
the depths for nearby cells at an angular distance of no more than 
$3^o$. As mean angular
deviation of muons during their passage through the rock, 
mainly caused by multiple Coulomb scattering, is about 0.5$^o$
(Antonioli et al., 1997), we excluded events if the slant depths
in the nearby cells were
less than 6.5 km. The values in the column $N_{7}^{cut}$ of Table 1 show 
the results after this cut. As a result of the depth cut we have 4 muon events 
against 1 background event which corresponds to the probability of 0.019.

During 1997 Markarian 501 had a remarkable flaring activity and was the 
brightest source in the sky at TeV energies. We used LVD data to look at 
possible enhancement of observed muons above calculated background during
the period of Mrk501 activity from the middle of March till the end of 
August, 1997. The results are presented in Table 2.


\begin{table}[htb]

\begin{center}
\begin{tabular}{|c|c|c|c|c|}
\hline
 Number of muons & all depths & $>$ 5 km w.e. & $>$ 7 km w.e. 
& $>$ 7 km w.e. after depth cut \\
\hline
Observed & 101 & 5 & 1 & 1\\
Simulated & 95 & 5.1 & 0.4 & 0.3\\
\hline
\end{tabular}
\end{center}
\end{table}   
\vspace{-1cm}
       
\section{Conclusions:}

We have confirmed a lack of serious systematic errors both in  
reconstruction of muon direction and pointing accuracy of the LVD detector by 
observing the effect of the Moon shadow with a statistical 
significance of 2.62 s.d.
Three depth (muon energy) ranges have been selected to search 
for point sources of VHE gammas.  
No statistically significant excess of muons above the simulated 
background has been found from any angular cell on the sky and 
for all energy ranges included in the 
analysis procedure. 
We have not found either any enhancement of muon flux
from the angular cells which include some known astrophysical sources.

\section{Acknowledgements:}

We wish to thank the staff of the Gran Sasso Laboratory
for their aid and collaboration. This work is supported by the
Italian Institute for Nuclear Phy\-sics (INFN) and in part by the
Italian Ministry of University and Scientific-Technological
Research (MURST), the Russian Ministry of Science and Technologies,
the US Department of Energy, the US National
Science Foundation, the State of Texas under its TATRP program,
and Brown University.
                                                     
\vspace{1cm}
\begin{center}
{\Large\bf References}
\end{center}
Aglietta, M. et al. 1992, Nuovo Cimento 105A, 1793.\\
Ahlen, S. et al., 1993, ApJ, 412, 301.\\
Antonioli, P. et al, 1997, Astrop. Phys., 7, 357. \\
Battistoni, G. et al., 1985, Phys. Lett. B., 155, 465.\\
Berezinsky, V.S. et al., 1988, A$\&$A, 189, 306.\\
Giglietto, N. et al., 1997, Proc. 25th ICRC (Durban), 6, 377.\\
Halzen, F., Stanev, T. and Yodh, G.B., 1997, Phys. Rev. D, 55, 4475.\\ 
Helene, O. 1983, Nucl. Instr. Meth., 212, 319.\\
Kudryavtsev, V.A., and Ryazhskaya, O.G., 1985, ZhETPh Lett.,42, 300.\\
Ong, R.A., 1998, Phys. Rep., 305, 93 and references therein.\\
Marshak, M. L. et al., 1985, Phys. Rev. Lett., 54, 2079.\\
Poirier, J. et al., 1977, Proc. 25th ICRC (Durban), 6, 370.\\ 
Stanev, T., Vankov, Ch. P., and Halzen, F., 1985, Proc. 17th ICRC (La Jolla),
7, 219.\\
Stanev, T., 1986, Phys. Rev. D, 33, 2740.\\
\end{document}